\documentclass{PoS}

\title{Recent progress of lattice and non-lattice super Yang-Mills }

\ShortTitle{Recent progress of lattice and non-lattice super Yang-Mills }

\author{\speaker{Masanori~Hanada} \\
Department of Physics, University of Washington, 
 Seattle, WA 98195-1560, USA\\
E-mail: \email{mhanada@u.washington.edu}}
\author{Issaku~Kanamori\\
Institut f\"ur Theoretische Physik, Universit\"at Regensburg, 
D-93040 Regensburg, Germany\\
E-mail: \email{issaku.kanamori@physik.uni-regensburg.de}}
\author{So~Matsuura\\
Department of Physics, and Research and Education Center for Natural Science, 
Keio University, 
4-1-1 Hiyoshi, Yokohama, 223-8521, Japan\\
E-mail: \email{s.matsu@phys-h.keio.ac.jp}}
\author{Fumihiko~Sugino\\
Okayama Institute for Quantum Physics, Kyoyama 1-9-1, Kita-ku, Okayama 700-0015, Japan\\
E-mail: \email{fumihiko.sugino@pref.okayama.lg.jp}}
\abstract{
We report recent progress of non-perturbative formulation of supersymmetric Yang-Mills. 
Although lattice formulations of two-dimensional theories which are fine tuning free 
to all order in perturbation theory are known for almost ten years, 
however, there were only few evidence for the validity at non-perturbative level. 
In this talk we argue that most numerical studies so far do not capture the physics in continuum, 
and add new evidence that lattice formulation works at non-perturbatively. 
We further point out that, by combining two-dimensional lattice and matrix model techniques inspired by D-brane dynamics in superstring theory, 
a non-perturbative formulation of the four-dimensional maximally supersymmetric Yang-Mills theory, 
which is free from the fine tuning at least to all order in perturbation theory, is obtained. }

\FullConference{The XXIX International Symposium on Lattice Field Theory\\
July 10-16, 2011\\
Squaw Valley, Lake Tahoe, California}

\begin{document}

\section{Introduction}
Supersymmetric Yang-Mills (SYM) theories play prominent roles in theoretical particle physics. 
In particular, maximally supersymmetric theories are expected to describe the non-perturbative nature of 
superstring/M theory \cite{BFSS,IKKT,MatrixString,Maldacena:1997re,Itzhaki:1998dd}. 
Lattice formulations of SYM theories are pursued because many important question would be answered 
only via numerical simulations.  
However, it is not a straightforward task, because the supersymmetry cannot be preserved completely on lattice: 
the SUSY algebra contains infinitesimal translation, 
which is broken on lattice by construction. Therefore, even if a given 
lattice theory converges to a supersymmetric theory at tree level, 
SUSY breaking corrections may arise radiatively. In order to control the divergence  
one needs some exact symmetries at discretized level. In 4d ${\cal N}=1$ pure SYM, by keeping 
the chiral symmetry one can obtain the correct supersymmetric continuum limit 
\cite{Kaplan:1983sk}\footnote{
For recent numerical studies, see \cite{Giedt:2008xm}. 
}. 
In several extended SYM theories, it is possible to keep a few supersymmetries unbroken. 
In two dimensions, those exact supersymmetries, together with other global symmetries, 
forbid SUSY breaking radiative corrections at least to all order in perturbation theory \cite{Kaplan:2002wv,CKKU03,Sugino04,2d other formulations}. 
(A similar statement holds for the Wess-Zumino model, and numerically tested in 
\cite{Catterall:2001fr}.)  

In one dimension (i.e. supersymmetric matrix quantum mechanics), 
the situation is much easier. Because the theory is UV finite, 
one does not have to rely on exact symmetries and hence 
a simple momentum cutoff prescription works \cite{Hanada:2007ti}. 
In fact, as demonstrated in \cite{Hanada:2007ti}, the momentum cutoff method is 
much more powerful than a usual lattice regularization and  
detailed Monte-Carlo studies have been done. 
In particular, the gauge/gravity duality between D0-brane quantum mechanics and type IIA superstring theory \cite{Itzhaki:1998dd} 
has been tested, and beautiful agreement including the $\alpha'$ correction has been confirmed 
\cite{Anagnostopoulos:2007fw,Catterall:2007fp}\footnote{
Other numerical studies in the context of the string theory can be found in 
\cite{Catterall:2010fx,Berenstein:2007wz,Campostrini:2004bs}, for example. 
}.   Furthermore long-disrance physics relevant for the matrix theory conjecture 
has also been studied \cite{Hanada:2009ne}. 

In order to study higher dimensions, one must establish fine-tuning free formulation of those theories. 
Obvious first step is non-perturbative test of two-dimensional theories. In particular one should see whether 
the fine-tuning-free nature persists to non-perturbative level. In \S~\ref{sec:2d} we show numerical results which strongly 
support the validity of these formulations at non-perturbative level. 
Based on the success of two-dimensional lattice formulation, in \S~\ref{sec:4d} we provide a hybrid formulation of 
two-dimensional lattice and matrix model techniques, which enables us to put four-dimensional maximally supersymmetric 
Yang-Mills theory on computer. 
\section{Non-perturbative test of two-dimensional lattice SYM}\label{sec:2d}
Numerical simulations of the supersymmetric gauge theories often suffer from the fermion sign problem\footnote{
In the case of maximally supersymmetric matrix quantum mechanics, 
agreement with the dual gravity prescription has been observed by ignoring the phase of the Pfaffian, 
even when the sign fluctuates violently \cite{Anagnostopoulos:2007fw}.  }. 
One of the exceptions is 4d ${\cal N}=1$ SYM and its dimensional reduction. Here we study 2d ${\cal N}=(2,2)$ SYM, 
which is the dimensional reduction to two dimensions. The action is 
\begin{eqnarray}
S
=
\frac{N}{\lambda}\int_0^{L_x} dx\int_0^{L_y}dy
\ Tr
\left\{
\frac{1}{4}F_{\mu\nu}^2
+
\frac{1}{2}(D_\mu X_i)^2
-
\frac{1}{4}[X_i,X_j]^2
-
\frac{1}{2}\bar{\psi}\Gamma^\mu D_\mu\psi
-
\frac{i}{2}\bar{\psi}\Gamma^i [X_i,\psi]
\right\}, 
\end{eqnarray}
where $\mu$ and $\nu$ run $x$ and $y$, $i$ and $j$ run $1$ and $2$, 
and $\Gamma^I=(\Gamma^\mu,\Gamma^i)$ are gamma matrices in four dimensions.   
$X_i$ are $N\times N$ hermitian matrices, $\psi_\alpha$ are $N\times N$ 
fermionic matrices with a Majorana index $\alpha$ 
and the covariant derivative 
is given by $D_\mu=\partial_\mu-i[A_\mu,\ \cdot\ ]$. 
The only parameters of the model are 
the size of circles $L_x$ and $L_y$. (Note that the coupling constant 
can be absorbed by redefining the fields and coordinates. 
Therefore we take the 't Hooft coupling $\lambda$ to be $1$.
Then the strong coupling corresponds to the large volume.) 

One obstacle for the simulation is the existence of the flat direction, 
along which two scalar fields $X_1$ and $X_2$ commute. 
In contrary to a theory on ${\mathbb R}^{1,3}$, there is no
superselection of the moduli parameter in this case. That is,
eigenvalues of scalars are determined dynamically.  
Therefore, some mechanism which restrict eigenvalues to a finite distribution 
is necessary for the stable simulation. 
In addition, to obtain an interesting dynamical system,
having a (small) finite region for the eigenvalues is important as well;
if the eigenvalues of the scalar spread so large, the theory would
run into the abelian phase, which is just a free theory\footnote{
Which phase is preferred is in fact a dynamical question.
At large-$N$, the flat direction is lifted and the system
stays non-abelian phase \cite{Hanada:2009hq,Hanada:2010qg}. 
This phase is an analogue of the black 1-brane solution in type IIB supergravity. 
}. 
In this work, we introduce soft SUSY-breaking mass to scalar fields 
\begin{equation}
 \mu^2 N\int d^2x\sum_{i=1,2} Tr X_i^2, 
\end{equation}
so that the flat direction is lifted. 
It is crucial to control the flat direction for various reasons. 
We have just mentioned two of them 
--- stability of the simulation and interesting non-abelian phase. 

In the Weyl notation, with an appropriate choice of the gamma matrices, the Dirac
operator is written as
\begin{eqnarray}
 D
\equiv
i\sigma^\mu D_\mu, 
\end{eqnarray}
where $\sigma^0=-i\textbf{1}_2$ and $\sigma^i(i=1,2,3)$ are Pauli matrices. 
By using $\sigma^2(i\sigma^\mu)\sigma^2=(i\sigma^\mu)^\ast$  
and the fact that $D_\mu$ is real in adjoint representation, we obtain 
\begin{eqnarray}
 \sigma^2 D\sigma^2
  =
  D^\ast.  
\end{eqnarray}
Therefore, if $\varphi$ is an eigenvector corresponding to an eigenvalue $\lambda$, 
$\sigma^2\varphi^\ast$ is also an eigenvector, with eigenvalue $\lambda^\ast$.
They are linearly independent and eigenvalues appear in a pair
$(\lambda,\lambda^\ast)$.
This assures the positivity of the determinant after removing $\lambda=0$ modes. 

\begin{figure}
 \hfil
 \includegraphics[width=0.49\linewidth]{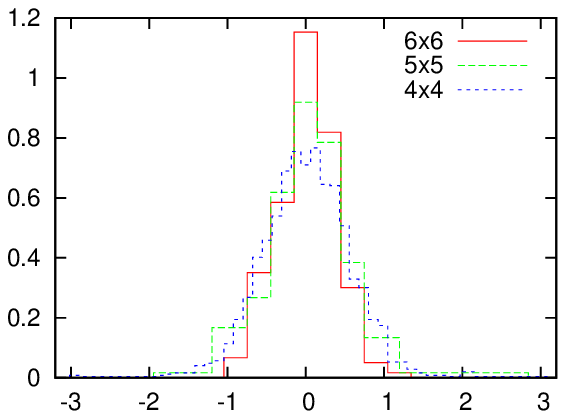}
 \hfil
 \includegraphics[width=0.49\linewidth]{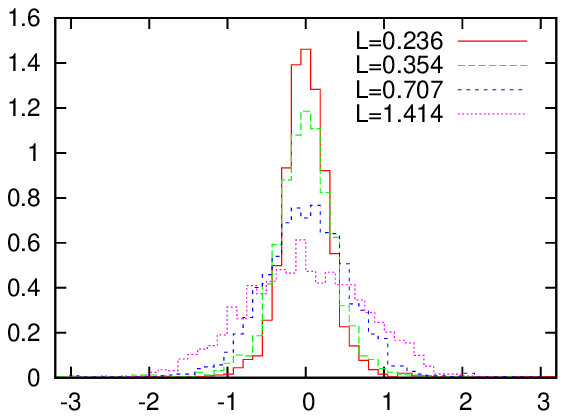}

  \caption{[Sugino] Argument of the Pfaffian in $SU(3)$ theory. 
The scalar mass is $\mu=0.20$.
 The left panel is for a fixed volume $0.707\times 0.707$
and thus different lattice spacings.
 The right panel is for a fixed $4\times 4$ lattice with various
 physical volumes (thus various lattice spacings).  (Figure from \cite{Hanada:2010qg})
}
 \label{fig:su3_pf}
\end{figure}

At discretized level, positivity of the determinant can be lost.  
In Cohen-Kaplan-Katz-Unsal (CKKU) model \cite{CKKU03} an existence of the sign problem had been reported. 
However it turned out that these studies did not capture the continuum physics, because the lattices were too coarse. 
In \cite{Hanada:2009hq} and \cite{Hanada:2010qg}, CKKU model and Sugino model \cite{Sugino04} have been studied 
and it has been confirmed that the determinant becomes real positive in the continuum limit (see also \cite{Kanamori:2009dk}). 
In Fig.~\ref{fig:su3_pf} we have plotted the distribution of the phase of the pfaffian\footnote{
Here we adopt the Majorana fermion. Hence we calculate the pfaffian rather than the determinant. 
} in the $SU(3)$ Sugino model. Here we chose $L_x=L_y=L$ and imposed periodic boundary condition for all fields along both directions. 
In the left panel we fixed the physical volume to be $L=0.707$ and used three lattice sizes, $4\times 4$, $5\times 5$ and $6\times 6$. 
It turns out the peak around zero becomes sharper as the lattice becomes finer. In the right panel we have fixed the lattice size and changed the lattice spacing. 
It can be seen that smaller lattice spacing gives sharper peak. Both plots clearly show that the sign disappears as the continuum limit is taken. 
For more details, see \cite{Hanada:2010qg}. 

\begin{figure}
 \hfil
 \includegraphics{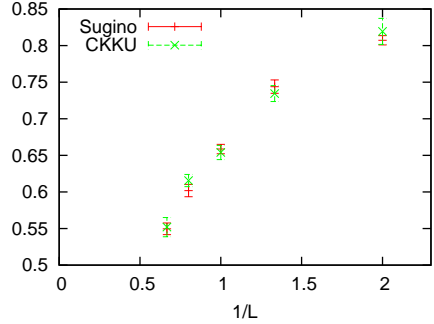}
  \caption{The expectation value of the Wilson loop
 $\langle|W|\rangle$ at $\mu=1.0$. Extrapolation to the
 continuum limit has been performed.
 The gauge grope is $U(2)$ and $SU(2)$, respectively. 
 (Figure from \cite{Hanada:2010qg})
}
 \label{fig:Wilson_comparison}
\end{figure}

Since we have established the absence of the sign problem in 2d ${\cal N}=(2,2)$ lattice SYM, we can perform various tests. 
So far, conservation of the supercurrent in the Sugino model has been observed in \cite{Kanamori:2008bk}. 
Here we add another result. 
In Fig.~\ref{fig:Wilson_comparison} we have plotted the expectation values of the Wilson loop winding on compactified circle, 
which are calculated in CKKU and Sugino models. Two models give the identical result, which strongly suggests that they converge 
to the same continuum limit without performing parameter fine tuning. 

Although theories with more supersymmetries suffer from the sign problem, it is a technical problem attached to the importance sampling 
and we can expect the restoration of all supersymmetries in the continuum limit also for them. 
Further investigation along this line is desirable.  
\section{Four-dimensional SYM out of two-dimensional lattice}\label{sec:4d}
In four dimensions, a few exact  supersymmetries are not strong enough to control the radiative correction, 
although it does reduce the number of fine tuning parameters down to three \cite{Catterall:2011pd}. 
In \cite{Hanada:2010kt} it has been pointed out that the fine tuning problem can be circumvented by 
using the matrix model techniques\footnote{
For another matrix model approach which works at large-$N$, see \cite{Ishii:2008ib}. 
}. 
The basic idea is simple. Consider the Berenstein-Maldacena-Nastase matrix model \cite{Berenstein:2002jq}, 
which is a deformation of the D0-brane matrix quantum mechanics keeping maximal supersymmetry. 
It has a fuzzy sphere solution, which is BPS. If one considers $k$-coincident fuzzy sphere, three-dimensional 
$U(k)$ non-commutative super Yang-Mills (NCSYM) on fuzzy sphere  is obtained. By taking appropriate large-$N$ limit,  
NCSYM on flat non-commutative plane is obtained, and by turning off the non-commutativity, one arrives at usual 
SYM on $R^{3}$ \cite{Maldacena:2002rb}. 
This method can easily be generalized to four-dimensional theory; if one starts with `BMN-like' two-dimensional theory 
with fuzzy sphere solutions, one can construct 4d theory. The crucial point is that such two-dimensional theory can be regularized by using the lattice. 
Bosonic part of the action of this two-dimensional theory is given by \cite{Hanada:2010kt} 
\begin{eqnarray}
S_{2d}
&=&
\frac{1}{g_{2d}^2}\int d^2x\ 
Tr\Bigl\{
\frac{1}{2}F_{12}^2
+
\frac{1}{2}(D_\mu X_I)^2
-
\frac{1}{4}[X_I,X_J]^2
+
\frac{\mu^2}{18}\sum_{a=1}^3 X_a^2
+
2i\mu X_1[X_2,X_3]
\nonumber\\ 
& &
\qquad
-
\frac{4\mu}{3} X_6[X_7,X_{8}] 
\Bigl\}. 
\label{continuum_bosonic}
\end{eqnarray}
The deformation by $\mu$ breaks 14 out of 16 SUSY softly. 
As shown in \cite{Hanada:2010kt}, this model can be regularized by lattice keeping two supercharges unbroken, 
and the continuum theory is obtained without parameter fine tuning, at least to all order in perturbation theory.

The continuum action (\ref{continuum_bosonic}) has constant BPS fuzzy sphere solution\footnote{
This background preserves exact supersymmetries at discretized level. 
} 
\begin{eqnarray}
X_a(x)=\frac{\mu}{3}L_a
\qquad(a=1,2,3), 
\qquad
X_i(x)=0
\qquad(i=4,\cdots,8), 
\end{eqnarray}
where $L_a$ are $M\times M$ matrices satisfying $SU(2)$ commutation relation 
\begin{eqnarray}
 [L_a,L_b]=i\epsilon_{abc}L_c. 
\end{eqnarray} 
By taking $k$-coincident fuzzy sphere solution, 
$L_a=L_a^{(M/k)}\otimes\textbf{1}_k$, where $L_a^{(M/k)}$ is the $(M/k)\times (M/k)$ 
irreducible representation, we obtain 4d $U(k)$ theory on fuzzy sphere.  
Essentially, adjoint action of $L_a$ is identified with the derivative 
and $[X_a,\ \cdot\ ]$ is regarded as the gauge covariant derivative 
\cite{Aoki:1999vr}. 
The noncommutativity is given by $\theta\sim k/(\mu^2 M)$ and UV/IR momentum cutoffs 
along spherical directions are $\mu M/k$ and $\mu$, respectively. 
4d coupling is given by $g_{4d}^2 = 2\pi\theta g_{2d}^2$. 
In order to get continuum 4d theory, we take large-$M$ and small $\mu$ limit while 
fixing $k$ and $g_{4d}^2$. 
In that limit, maximal supersymmetry is restored 
because soft SUSY breaking parameter $\mu$ goes to zero. 
One can take a limit with any value of non-commutativity $\theta$, 
and $\theta\to 0$ limit is expected to be smooth in the maximally supersymmetric theory. 
That the limit should be smooth is natural physically, because a possible obstacle 
is a new IR divergence arising due to the UV/IR mixing reflecting the UV divergence, 
which should be absent in UV finite theories. 

In the above we assumed the radius of the fuzzy sphere does not deviate from classical value. 
Whether it is the case or not should be tested by numerics. If the radius is renormalized, 
we should take it into account by replacing the parameters in the mapping rule with 
renormalized ones. 
\section{Conclusion}
We reported recent progress of the non-perturbative formulations of SYM. 
Lattice formulations of two-dimensional ${\cal N}=(2,2)$ SYM has been tested non-perturbatively  
and confirmed to work without parameter fine tuning beyond the perturbative level. 
It is also shown that four-dimensional SYM can be constructed by combining two-dimensional lattice and 
a matrix model method. 

There are various problems one can attack based on above results. 
Firstly, 2d ${\cal N}=(2,2)$ SYM can have interesting physics on its own 
(see e.g. \cite{Kanamori:2009dk,Hori:2006dk,Kanamori:2007yx,Kanamori:2008yy,Dorigoni:2010jv}) and further numerical study is desirable. 
2d maximal SYM is even more interesting because it describes \cite{AMMW} the black hole/black string phase transition \cite{GrLa}.   
It is important to establish the absence of fine tuning in this case and perform a large-scale simulation along the line of \cite{Catterall:2010fx}. 
Numerical study of four-dimensional SYM would be the most important. 
The formulation explained in \S~\ref{sec:4d} would be useful to understand the AdS/CFT duality further, especially 
away from the weak and strong coupling. Another interesting class of theories are SUSY QCD.  
At large-$N$,  the large-$N$ reduction \cite{Ishii:2008ib} can be applied to SUSY QCD \cite{Hanada:2009hd}. 
It would provide us with a tool to study important problems e.g. the study of spontaneous SUSY breakdown or the test of Seiberg duality. 
It is interesting especially because it may have a close connection to physics at LHC.   
\section*{Acknowledgments}
The work of M.~H. is supported by Japan Society for the Promotion of Science Postdoctoral Fellowships for Research Abroad.  
I. K. is supported by the EU ITN STRONGnet and the DFG SFB/Transregio 55. 
F. S. is supported in part by Grant-in-Aid for Scientific Research (C), 21540290. 


\end{document}